\documentclass[conference]{IEEEtran}
\IEEEoverridecommandlockouts
\usepackage{cite}
\usepackage{amsmath,amssymb,amsfonts}
\usepackage{algorithmic}
\usepackage{graphicx}
\usepackage{textcomp}
\usepackage{xcolor}
\usepackage{hyperref}
\usepackage{multirow}

\def\BibTeX{{\rm B\kern-.05em{\sc i\kern-.025em b}\kern-.08em
    T\kern-.1667em\lower.7ex\hbox{E}\kern-.125emX}}

\begin{document}

\title{Quality-Controlled Multimodal Emotion Recognition in Conversations with Identity-Based Transfer Learning and MAMBA Fusion\ \\ \ \\
  \small{Technical Report: \href{https://www.recognitiontechnologies.com/~beigi/ps/RTI-20251118-01.pdf}{RTI-20251118-01}}\\
  \small{\href{http://dx.doi.org/10.13140/RG.2.2.33632.55045}{DOI: 10.13140/RG.2.2.33632.55045}}}

\author{
\IEEEauthorblockN{Zanxu Wang$^{1}$, Homayoon Beigi$^{1,2}$}
\IEEEauthorblockA{$^{1}$Columbia University, New York, USA}
\IEEEauthorblockA{$^{2}$Recognition Technologies, Inc., New York, USA}
\IEEEauthorblockA{$^{1}$zw2864@columbia.edu, $^{2}$beigi@recotechnologies.com}
}

\maketitle

\begin{abstract}
This paper addresses data quality issues in multimodal emotion recognition in conversation (MERC) through systematic quality control and multi-stage transfer learning. We implement a quality control pipeline for MELD and IEMOCAP datasets that validates speaker identity, audio-text alignment, and face detection. We leverage transfer learning from speaker and face recognition, assuming that identity-discriminative embeddings capture not only stable acoustic and facial traits but also person-specific patterns of emotional expression. We employ RecoMadeEasy\textsuperscript{\textregistered} engines for extracting 512-dimensional speaker and face embeddings, fine-tune MPNet-v2 for emotion-aware text representations, and adapt these features through emotion-specific MLPs trained on unimodal datasets. MAMBA-based trimodal fusion achieves 64.8\% accuracy on MELD and 74.3\% on IEMOCAP. These results show that combining identity-based audio and visual embeddings with emotion-tuned text representations on a quality-controlled subset of data yields consistent competitive performance for multimodal emotion recognition in conversation and provides a basis for further improvement on challenging, low-frequency emotion classes.
\end{abstract}


\section{Introduction}
Multimodal emotion recognition in conversation (MERC) leverages complementary information from text, audio, and visual modalities to identify emotional states in dialogue contexts \cite{r-m:pantic-2005,r-m:ananthram-2020}. While recent architectures demonstrate incremental improvements in MERC performance, such as MGCMA \cite{r-m:wang-2025} and MMGAT-EMO \cite{r-m:zhang-2025}, fundamental data quality issues in benchmark datasets remain unaddressed.

Large-scale conversational emotion datasets such as MELD \cite{r-m:poria-2019} serve as primary benchmarks for MERC research. However, these datasets exhibit systematic quality issues that propagate through modeling pipelines: speaker labels conflate distinct individuals (e.g., ``Waiter'' assigned to multiple actors), audio-text misalignment occurs due to imprecise video segmentation, and face tracking often captures non-speaking individuals or fails entirely. Despite these issues affecting a substantial portion of samples, few studies systematically address data curation.

Additionally, beyond data quality, most MERC approaches rely on generic pre-trained models for each modality that lack emotion-specific discriminability. While numerous unimodal emotion datasets exist with prototypical emotional expressions—acted speech in RAVDESS \cite{r-m:livingstone-2018}, posed facial expressions in CK+ \cite{r-m:lucey-2010}, emotion-labeled text in GoEmotions \cite{r-m:demszky-2020}—their potential to inform multimodal models through transfer learning remains largely unexplored.

The selection of appropriate pre-trained models for transfer learning is critical. We leverage the RecoMadeEasy\textsuperscript{\textregistered} speaker and face recognition models built by Recognition Technologies, Inc.~\cite{r-m:recotech}, based on the hypothesis that identity-preserving representations inherently capture fine-grained acoustic and facial patterns that correlate with emotional expression. Speaker embeddings~\cite{r-m:beigi-book-2011} encode prosodic variations, voice quality, and temporal dynamics—all crucial for emotion recognition. Similarly, face recognition models must learn subtle facial muscle movements and expression dynamics to distinguish individuals, providing rich features for emotion classification.

Finally, fusion strategies predominantly employ attention-based cross-modal transformers \cite{r-m:siriwardhana-2020,r-m:rahman-2020} with quadratic complexity, raising questions about efficiency in practical settings. Recent efficient sequence models like MAMBA \cite{r-m:gu-2023} offer linear complexity but have not been widely used for multimodal emotion fusion.

This paper addresses these challenges through an integrated methodology with three core contributions:

\textbf{1. Systematic Data Quality Control:} We develop a systematic quality control pipeline for MELD and IEMOCAP that resolves speaker disambiguation, audio-text alignment verification, and face-speaker validation.

\textbf{2. Emotion-Specific Multi-Stage Transfer Learning:} We propose leveraging various auxiliary unimodal emotion datasets to inject emotion-discriminative information into each modality's representation before fusion. This multi-stage approach uses emotion-specific text datasets to fine-tune language models, vocal emotion corpora to adapt speaker embeddings, and facial expression databases to condition face features.

\textbf{3. Efficient Fusion with State Space Model:} We demonstrate that a single MAMBA block achieves competitive performance with linear complexity, and report empirical results on quality-controlled data subsets while identifying remaining challenges for future investigation.

\begin{figure*}[!t]
\centering
\includegraphics[width=\textwidth]{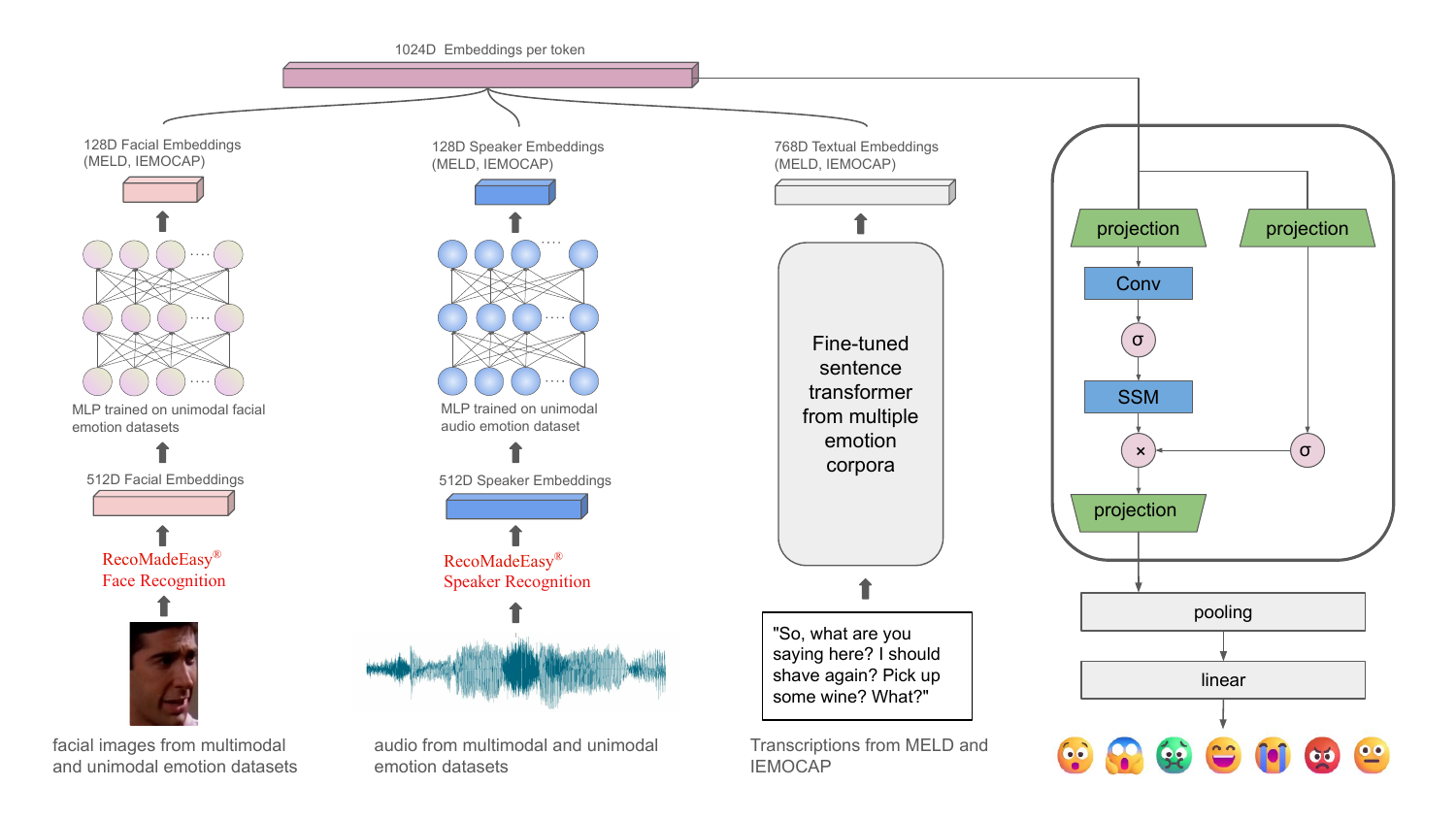}
\caption{Overview of the proposed three-stage quality-controlled multimodal emotion recognition pipeline. Stage~1 extracts foundation embeddings from RecoMadeEasy\textsuperscript{\textregistered} speaker and face recognition engines~\cite{r-m:recotech} and a fine-tuned MPNet-based sentence transformer for MELD and IEMOCAP utterances. Stage~2 adapts the 512-dimensional audio and visual features into 128-dimensional emotion embeddings using modality-specific MLPs trained on unimodal emotion datasets. Stage~3 fuses 768-dimensional text, 128-dimensional speaker, and 128-dimensional facial embeddings into 1024-dimensional token representations and applies a single MAMBA block, pooling, and a linear classification head for utterance-level emotion prediction.}
\label{fig:pipeline}
\end{figure*}

\section{Related Work}

\subsection{Transfer Learning in Speech Emotion Recognition}
Transfer learning has emerged as a dominant paradigm for addressing data scarcity in emotion recognition. Recent work by Jakubec et al.~\cite{r-m:jacubec-2024} demonstrates that speaker embedding models (d-vector, x-vector, r-vector) pre-trained on large speaker verification~\cite{r-m:beigi-book-2011} datasets significantly improve speech emotion recognition (SER). The advantage lies in these models' ability to capture speaker-specific acoustic patterns including prosodic variations and voice quality characteristics that correlate with emotional states \cite{r-m:padi-2022}.

Padi et al.~\cite{r-m:padi-2022} proposed leveraging ResNet-based models trained on speaker recognition~\cite{r-m:beigi-book-2011} tasks, incorporating statistics pooling layers to handle variable-length inputs. Their approach eliminates the need for sequence truncation commonly used in SER systems, achieving competitive results on IEMOCAP. This supports our hypothesis that speaker identity models capture emotion-relevant acoustic features.

\subsection{Face Recognition for Emotion Transfer}
The relationship between face recognition and emotion recognition has been extensively studied. Knyazev et al.~\cite{r-m:knyazev-2017} demonstrated that industry-level face recognition networks pre-trained on large-scale datasets (e.g., VGG-Face2) outperform emotion-specific models when fine-tuned for facial emotion recognition (FER). This finding suggests that learning to distinguish facial identities requires capturing subtle muscular movements and expression dynamics that directly transfer to emotion recognition.

Ngo and Yoon \cite{r-m:ngo-2020} showed that an SE-ResNet-50 model pre-trained on the VGG-Face2 database, combined with a novel weighted-cluster loss function, effectively transfers high-level facial features to FER tasks. The success of face recognition models in emotion tasks stems from their need to be invariant to expressions while still encoding them—creating representations that separate identity from emotion while preserving both.

\subsection{Sentence Transformers for Emotion-Aware Text}
While BERT and its variants dominate text emotion recognition, recent advances in sentence transformers offer superior semantic representations. MPNet \cite{r-m:song-2020} combines the strengths of BERT's bidirectional context and XLNet's autoregressive modeling, addressing the limitations of masked language modeling through permuted language modeling. The all-mpnet-base-v2 model, fine-tuned on over 1 billion sentence pairs using contrastive learning \cite{r-m:reimers-2019}, produces 768-dimensional embeddings that capture nuanced semantic relationships.

The choice of MPNet-v2 over alternatives is motivated by several factors: (1) the contrastive training objective naturally aligns with emotion recognition's need to distinguish subtle semantic differences at utterance level; (2) extensive evaluation shows MPNet-based models achieve state-of-the-art performance on semantic textual similarity tasks \cite{r-m:reimers-2019,r-m:reimers-2021}, which correlate with emotion understanding.

\subsection{Quality Issues in MERC Datasets}
Despite widespread use of MELD and IEMOCAP, few studies systematically address their quality issues. Busso et al.~\cite{r-m:busso-2008} acknowledged annotation challenges in the original IEMOCAP paper, noting inter-annotator agreement issues for ambiguous emotions. Recent work highlights that fear and disgust categories in MELD barely exceed chance-level recognition even with state-of-the-art models \cite{r-m:meng-2024}, suggesting the possibility of fundamental data problems rather than model limitations.

The lack of systematic quality control might lead to inflated performance claims. Models trained on corrupted data may learn spurious correlations by using audio-text misalignment as unintended features. Our work aims to narrow this gap through comprehensive validation of modality alignment and speaker identity.

\section{Methodology}

\subsection{Problem Formulation}
Given a conversational sequence $U = [u_1, u_2, \ldots, u_N]$ with $N$ utterances and corresponding speakers $S = [s_1, s_2, \ldots, s_N]$, each utterance $u_i$ comprises three modalities: text $t_i$, audio $a_i$, and visual (face) $v_i$. The goal is to predict emotion label $y_i \in \mathcal{Y}$ for each utterance, where $|\mathcal{Y}| = 7$ for MELD and $|\mathcal{Y}| = 4$ for IEMOCAP.

\subsection{Multi-Stage Emotion Representation Learning}
Our proposed framework (Figure~\ref{fig:pipeline}) consists of a hierarchical multimodal pipeline across three stages.

\noindent\textbf{1. Foundation Embedding Extraction}

We begin by extracting unimodal embeddings from pretrained speaker recognition and face recognition models, as well as a fine-tuned MPNet-v2 model, as foundation embeddings for each modality.

\noindent a) \textbf{Speaker Embeddings}

We leverage the RecoMadeEasy\textsuperscript{\textregistered} speaker recognition engine~\cite{r-m:recotech} to extract embeddings from audio. Specifically, we pass both unimodal audio datasets (e.g., CREMA-D and RAVDESS) and audio from MELD and IEMOCAP to the engine, obtaining 512-dimensional speaker embeddings per utterance.

\noindent b) \textbf{Face Embeddings}

For visual features, extracted facial images from MELD, IEMOCAP, and additional unimodal facial expression datasets (e.g., CK+ and RAF-DB) are processed through the RecoMadeEasy\textsuperscript{\textregistered} face recognition engine to obtain 512-dimensional face embeddings.

\noindent c) \textbf{Textual Embeddings}

Textual embeddings are derived from the \texttt{sentence-transformers/all-mpnet-base-v2} model, which we fine-tune for seven-class emotion classification (surprise, fear, disgust, happiness, sadness, anger, neutral) using a supervised learning setup with balanced sampling from multiple emotion corpora. A classification head with two linear layers is added on top of the pretrained MPNet encoder, with the final layer containing seven neurons, one for each emotion class. The input text is tokenized using the MPNet tokenizer with padding to a maximum length and truncation enabled. Training is performed for three epochs using the AdamW optimizer with a learning rate of $2\times10^{-5}$, a batch size of 16, and weight decay of $0.01$. We save the best checkpoint based on validation loss at the end of each epoch. This fine-tuning procedure yields a training accuracy of 90.08\% and a test accuracy of 83.41\% on the aggregated emotion corpora. After training, we discard the classification head and use the 768-dimensional pooled output of the encoder as the textual embedding for each utterance. The transcripts of IEMOCAP and MELD dialogues are passed through this fine-tuned MPNet model to produce 768-dimensional textual embeddings that encode both semantic and emotional content.

\noindent\textbf{2. Modality-Specific Adaptation}

Once we obtain the initial 512-dimensional embeddings for the audio (speaker) and facial (image) modalities from the Stage~1 feature extraction process, we focus on adapting these embeddings to better reflect emotional content. Our goal is to transform the raw modality-specific features into representations that are sensitive to emotional patterns.

Dedicated unimodal emotion datasets contain clearer, more prototypical expressions of each emotion (for example, exaggerated vocal tones in RAVDESS and posed facial expressions in CK+) compared to natural conversational data, where emotions are often subtle or mixed. This transfer learning strategy teaches the model what each emotion looks and sounds like before applying it to noisy multimodal conversational context.

To achieve this, we train feedforward multi-layer perceptron (MLP) classifiers on unimodal emotion recognition datasets, each containing single-modality emotional cues. For facial expression recognition, we train an MLP on 512-dimensional face embeddings extracted from static facial images for seven-class emotion classification. The MLP architecture consists of three fully connected layers with dimensions $512 \rightarrow 256 \rightarrow 128 \rightarrow 7$, with batch normalization and ReLU activation after each hidden layer. Dropout with a rate of $0.3$ is applied after each hidden layer to reduce overfitting. The model is trained only on non-MELD datasets (including CK+ and RAF-DB) using the Adam optimizer with an initial learning rate of $1\times10^{-3}$ and weight decay of $1\times10^{-5}$. The learning rate is adjusted by a ReduceLROnPlateau scheduler with a factor of $0.1$ and patience of 5 epochs. Training uses cross-entropy loss with a batch size of 32 for up to 100 epochs, with early stopping triggered after 10 epochs without improvement in validation F1-score. This setup achieves about $82.5\%$ validation accuracy on the non-MELD facial expression test set. 

We apply the same MLP architecture, hyperparameters, and training procedure to 512-dimensional audio embeddings derived from emotion speech datasets (CREMA-D, RAVDESS, SAVEE, and TESS). This audio MLP reaches $69\%$ validation accuracy on the non-MELD audio test sets.

For both audio and face, we use the 128-dimensional penultimate layer of the MLP as the transformed emotion embedding. These 128-dimensional vectors serve as the final unimodal representations used in the subsequent multimodal fusion stage, as they capture the emotion-relevant structure learned from unimodal data while remaining compact.

\noindent\textbf{3. Multimodal Fusion for Classification}

We employ a single MAMBA block ($d_{\text{state}}=64$, $\text{expand\_factor}=2$, $d_{\text{conv}}=4$) followed by mean pooling and a linear classification head for utterance-level emotion recognition. The model's input dimension ($d_{\text{model}}$) varies based on the modality combination: 1024-dimensional for text+speaker+face, 768-dimensional for text-only, and other configurations for different multimodal setups. Each input utterance is represented as a variable-length sequence of concatenated multimodal embeddings, with padding and attention masking applied to handle sequences of different lengths. We train on MELD and IEMOCAP training sets separately using AdamW optimizer with a learning rate of $1 \times 10^{-5}$, batch size 32, and early stopping (patience 10 epochs) based on validation weighted F1. For MELD (7 emotion classes), we address severe class imbalance using weighted cross-entropy loss with class weights of [15.0, 15.0, 6.0, 1.0, 3.0, 6.0, 4.0] for surprise, fear, disgust, happiness, sadness, anger, and neutral respectively, combined with label smoothing of 0.2. For IEMOCAP (4 emotion classes with excitement merged into happiness), we use standard cross-entropy loss with the same label smoothing of 0.2.

\noindent a) \textbf{Text-only:} We feed each token's 768-dimensional embedding sequentially into the model.

\noindent b) \textbf{Face-only:} We feed each video frame's 128-dimensional reduced feature vector into the model.

\noindent c) \textbf{Text + Face Fusion:} For a given utterance, we first determine the correspondence between text tokens and video frames. With 15 tokens and 4 frames, we match the first 4 tokens to frame 1, the next 4 to frame 2, the next 4 to frame 3, and the final 3 to frame 4. Each token's 768-dimensional embedding is concatenated with its corresponding frame's 128-dimensional face representation, yielding an 896-dimensional vector per token.

\noindent d) \textbf{Speaker Embedding Integration:} When including speaker information, we concatenate the same speaker embedding vector for the entire utterance alongside the text and/or face features. This maintains a consistent speaker representation within the fused input sequence.

\subsection{Datasets}
\begin{figure}[!htbp]
\centering
\includegraphics[width=\columnwidth]{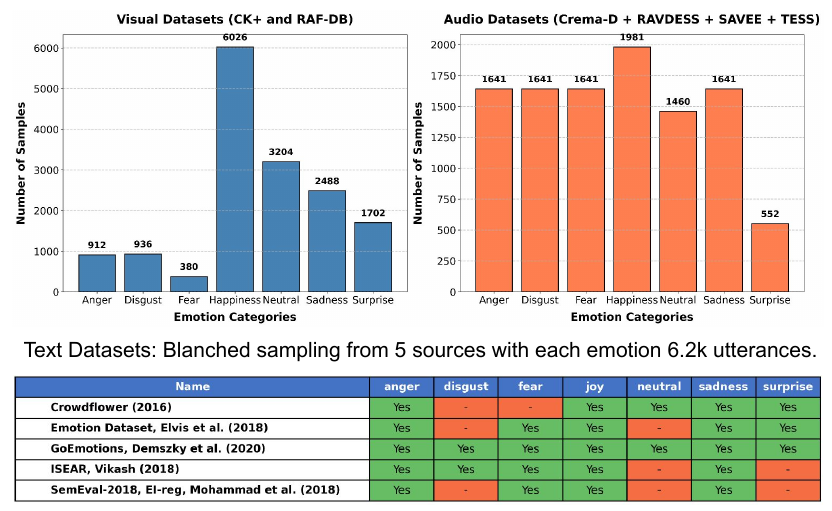}
\caption{Distribution of emotion categories across auxiliary unimodal datasets. \textbf{Top left:} Visual datasets (CK+ and RAF-DB) showing 15,648 total samples. \textbf{Top right:} Audio datasets (CREMA-D, RAVDESS, SAVEE, and TESS) with 10,557 samples. \textbf{Bottom:} Text dataset composition showing balanced sampling from five sources (CrowdFlower, CARER, GoEmotions, ISEAR, and SemEval-2018) totaling 6.2k utterances across seven emotion categories.}
\label{fig:unimodal_distribution}
\end{figure}

We employ a comprehensive collection of multimodal and unimodal emotion datasets for our experiments. The multimodal datasets (MELD and IEMOCAP) serve as our target evaluation benchmarks, while the unimodal datasets are used for fine-tuning MPNet-v2 and training the face and speaker MLPs in Stage 2 of our framework. Figure~\ref{fig:unimodal_distribution} illustrates the distribution of emotion categories across all auxiliary unimodal datasets, demonstrating our balanced sampling strategy across modalities.

\begin{figure*}[!t]
\centering
\includegraphics[width=\textwidth]{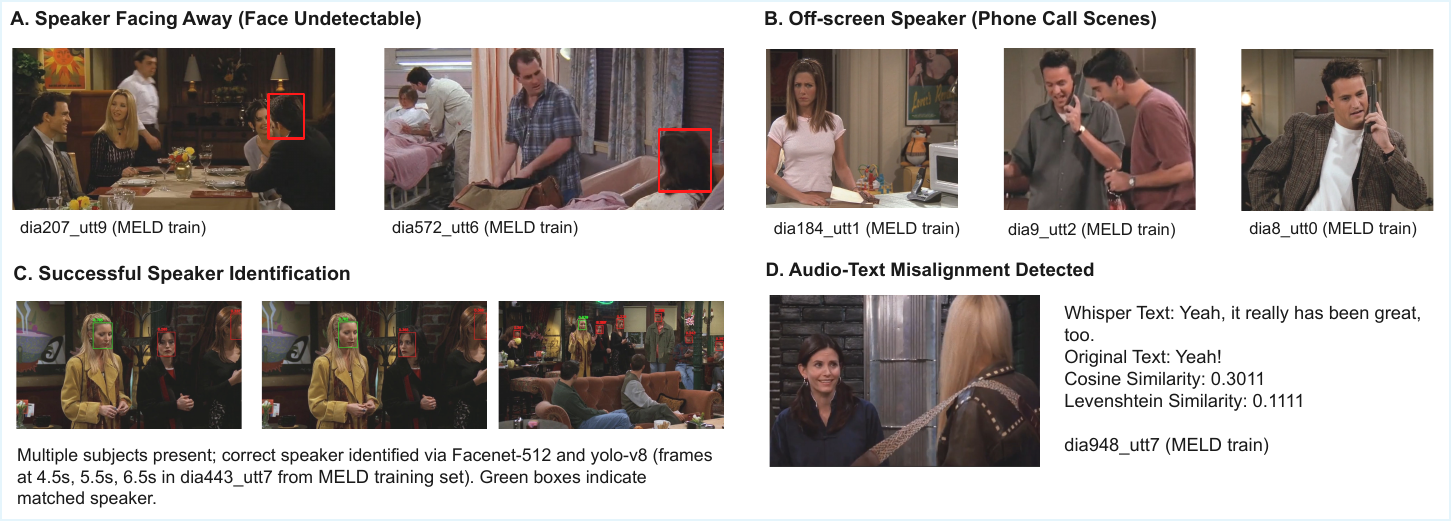}
\caption{Examples of quality control challenges and solutions in MELD dataset processing. \textbf{(A) Speaker Facing Away:} Utterances where the speaker's face is not visible (red box indicates detection failure) are removed to ensure reliable facial expression analysis. \textbf{(B) Off-screen Speaker:} Phone call scenes where the speaking character is not visible on camera are filtered out. \textbf{(C) Successful Speaker Identification:} Multi-person scenes where YOLOv8 detects multiple faces and Facenet-512 correctly identifies the target speaker (green boxes) based on embedding similarity across temporal frames (4.5s, 5.5s, 6.5s). \textbf{(D) Audio-Text Misalignment:} Example of detected misalignment between original transcript (``Yeah!'') and Whisper transcription (``Yeah, it really has been great, too.''), identified through low cosine similarity (0.30) and Levenshtein similarity (0.11) scores.}
\label{fig:pipeline_qc}
\end{figure*}

\noindent\textbf{1. Target Conversational Datasets}

\textbf{MELD (Multimodal EmotionLines Dataset)} \cite{r-m:poria-2019} is a large-scale multimodal dataset derived from TV show dialogues, providing synchronized audio, video and text data for more than 13,000 utterances in 1,400 dialogues. It includes seven emotion categories: anger, disgust, fear, joy, neutral, sadness, and surprise. The dataset's focus on conversational context makes it particularly suitable for emotion recognition in multi-turn dialogues, where emotions vary dynamically across speakers and modalities.

\textbf{IEMOCAP (Interactive Emotional Dyadic Motion Capture Database)} \cite{r-m:busso-2008} is a widely used multimodal dataset containing approximately 12 hours of audio, video, and text from both scripted and improvised dyadic conversations between actors. It encompasses nine emotion categories: angry, fearful, disappointed, happy, sad, surprised, excited, frustrated, and neutral. The dataset provides detailed facial expressions, vocal signals, and transcriptions, making it ideal for comprehensive multimodal emotion analysis.

\noindent\textbf{2. Auxiliary Unimodal Emotion Datasets}

To enhance modality-specific representations, we leverage several established unimodal emotion datasets across visual, audio, and text modalities. As shown in Figure~\ref{fig:unimodal_distribution}, we apply balanced sampling strategies to ensure comparable representation across emotion categories within each modality.

\noindent a) \textbf{Facial Expression Datasets}

\textbf{CK+ (Extended Cohn-Kanade Dataset)} \cite{r-m:lucey-2010} contains 593 video sequences from 123 subjects displaying posed facial expressions across seven emotions. Each sequence begins with a neutral expression and progressively intensifies to a peak expression, providing valuable data for expression dynamics.

\textbf{RAF-DB (Real-World Affective Faces Database)} \cite{r-m:li-2017} is a large-scale dataset comprising approximately 30,000 crowd-labeled facial images covering both basic emotions (anger, disgust, fear, happiness, sadness, surprise, neutral) and compound emotions (e.g., happily surprised, sadly fearful). Its real-world collection conditions enhance model generalization.

\noindent b) \textbf{Vocal Expression Datasets}

\textbf{CREMA-D (Crowd-sourced Emotional Multimodal Actors Dataset)} \cite{r-m:cao-2014} includes over 7,000 audio-visual recordings from 91 actors expressing six emotions: anger, disgust, fear, happiness, neutral, and sadness. The dataset's diversity in actor demographics provides robust vocal expression patterns.

\textbf{RAVDESS (Ryerson Audio-Visual Database of Emotional Speech and Song)} \cite{r-m:livingstone-2018} contains recordings from 24 professional actors producing both speech and song across eight emotions, including calm, happy, sad, angry, fearful, surprised, and disgusted, with two intensity levels (normal and strong).

\textbf{TESS (Toronto Emotional Speech Set)} \cite{r-m:dupuis-2010} comprises 2,800 audio clips from two female speakers aged 26 and 64, each expressing seven emotions. The age diversity makes it particularly valuable for studying vocal emotion cue variations across different age groups.

\textbf{SAVEE (Surrey Audio-Visual Expressed Emotion Database)} \cite{r-m:haq-2008} includes recordings of four male actors expressing seven emotions: anger, disgust, fear, happiness, sadness, surprise, and neutral. The dataset provides high-quality audio recordings suitable for acoustic feature extraction.

\noindent c) \textbf{Textual Emotion Datasets}

\textbf{CrowdFlower (Emotion in Text Dataset, 2016)} is a corpus of 40,000 tweets labeled with 13 emotion categories (e.g., anger, joy, sadness, fear, enthusiasm, etc.) via crowd-sourcing. We map these labels into our seven-class taxonomy and sample 6.2k utterances to balance with other sources.

\textbf{CARER (Contextual Affect Dataset)} \cite{r-m:saravia-2018} covers five basic emotions (anger, fear, joy, sadness, and surprise) with sentences collected from diverse domains, providing varied linguistic expressions of emotion.

\textbf{GoEmotions} \cite{r-m:demszky-2020} is a large-scale dataset of 58,000 Reddit comments labeled with 27 fine-grained emotion categories. We map its labels to our seven-emotion taxonomy, leveraging its rich emotional diversity and conversational context.

\textbf{ISEAR (International Survey on Emotion Antecedents and Reactions)} \cite{r-m:scherer-1994} contains self-reported emotional experiences across multiple emotions (anger, disgust, fear, joy, sadness) from participants in 37 countries. The dataset includes contextual descriptions of emotion triggers, providing valuable semantic information.

\textbf{SemEval-2018 Task 1 (Affect in Tweets)} \cite{r-m:mohammad-2018} provides emotion intensity ratings for tweets in four emotions (anger, fear, joy, sadness) and multi-label classification for 11 emotions. We discretize the intensity ratings into categorical labels and select tweets covering our seven-emotion set.

We standardize all datasets to a unified seven-emotion taxonomy (anger, disgust, fear, joy/happiness, neutral, sadness, surprise) where applicable. As illustrated in Figure~\ref{fig:unimodal_distribution}, our text modality employs balanced sampling from five sources, resulting in approximately 6.2k utterances per emotion. The visual datasets (CK+ and RAF-DB) provide 15,648 samples, while the audio datasets (CREMA-D, RAVDESS, TESS, and SAVEE) contribute 10,557 samples.

\subsection{Data Quality Control Pipeline}
Our quality control pipeline ensures the integrity, quality, and modality alignment of data from the MELD and IEMOCAP datasets. Figure~\ref{fig:pipeline_qc} illustrates key challenges and solutions in our data cleaning process, including speaker identification in multi-person scenes, handling off-screen speakers, and detecting audio-text misalignments. The objective is to generate a clean, speaker-specific, and temporally aligned corpus of audiovisual utterances suitable for emotion recognition.

\noindent\textbf{1. Face Extraction from MELD.}

\noindent a) \textbf{Database Construction for Unique Speakers:} MELD contains over 300 distinct speakers, but in many cases different individuals share identical speaker labels. To address this ambiguity, we created a speaker database with corrected annotations:

\begin{itemize}
\item We re-labeled speakers who shared the same name by appending a dialogue-specific ID, resulting in 352 uniquely identified speakers.
\item Entries with multiple speakers (e.g., ``All,'' ``Guys,'' or ``Phoebe and Rachel'') were removed to ensure that each utterance corresponded to a single speaker.
\end{itemize}

\noindent b) \textbf{Sampling and Face Detection:} For each of the 352 speakers, we sampled up to 15 video clips. We used the DeepFace library with a YOLOv8 backend to detect faces as follows:
\begin{itemize}
\item Extract one frame per second, starting from 0.5 seconds into the utterance.
\item Save every detected face in each frame to the corresponding speaker's folder.
\end{itemize}

All cropped face images were inspected to remove incorrect detections and misaligned utterances (Figure~\ref{fig:pipeline_qc}A-B):
\begin{itemize}
\item Utterances where the speaker never appeared (e.g., speaker off-screen or on the phone) were discarded.
\item Utterances where the video and the transcription were not aligned were removed.
\end{itemize}

From these steps, we eliminated 182 utterances.

\noindent c) \textbf{Ensuring Audio-Text Alignment in MELD:} MELD's original video cropping sometimes resulted in misalignments between audio and text, especially for short utterances (1--2 seconds long). To address this, we re-transcribed the audio using the Whisper API and compared the newly generated transcriptions to the original MELD annotations. Ninety-one utterances were removed due to extremely short or inaudible audio, where Whisper yielded empty transcriptions. To detect more subtle misalignments (Figure~\ref{fig:pipeline_qc}D), we employed two metrics:
\begin{itemize}
\item \textbf{Cosine Similarity (Textual Embeddings):} Identified utterances with semantically different transcriptions.
\item \textbf{Levenshtein Similarity:} Captured utterances that differed significantly in length, indicating misalignment.
\end{itemize}

\begin{figure}[htbp]
\centering
\includegraphics[width=\columnwidth]{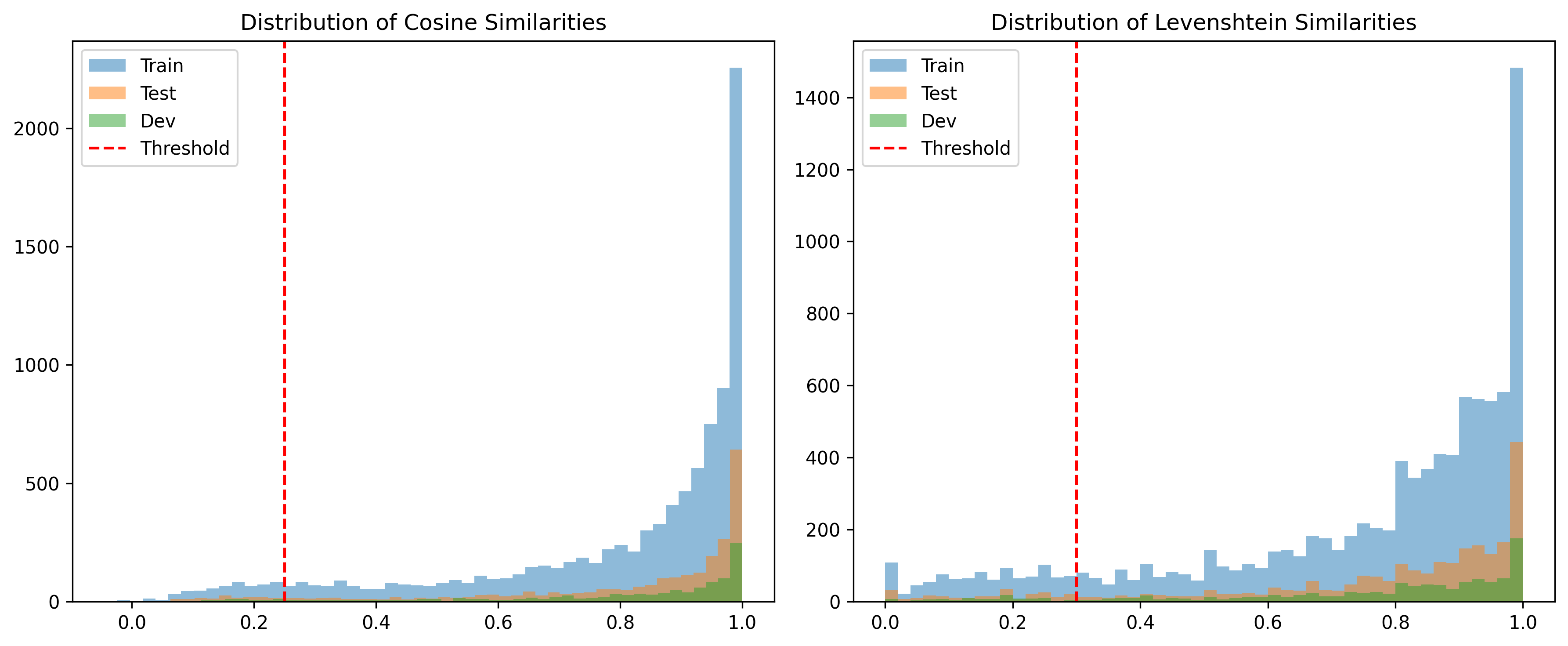}
\caption{Distribution of audio-text alignment metrics across MELD splits. \textbf{Left:} Cosine similarity between original and Whisper-generated transcriptions using MPNet-v2 embeddings. \textbf{Right:} Levenshtein similarity measuring character-level differences. Red dashed lines indicate filtering thresholds (0.25 for cosine similarity, 0.3 for Levenshtein). Most utterances show high alignment, while removed samples cluster near low similarity scores.}
\label{fig:similarity}
\end{figure}

We passed each utterance's Whisper transcription and original transcription to the MPNet-v2 model, extracted the sentence embeddings, and calculated cosine similarity scores. As shown in Figure~\ref{fig:similarity}, we set a threshold of 0.25 for cosine similarity and 0.3 for normalized Levenshtein similarity. Only utterances passing both thresholds were retained. The distribution reveals that most utterances exhibit high alignment (cosine similarity > 0.9), while the filtering process effectively removes outliers with significant mismatches.

\noindent d) \textbf{Face Extraction Based on Speaker Identity:} We used the Facenet-512 model within DeepFace to generate embeddings for each speaker's sample images. By averaging these embeddings, we obtained a unique 512-dimensional representation per speaker. For each utterance, we re-detected faces using YOLOv8 and compared them to the speaker's stored embedding (Figure~\ref{fig:pipeline_qc}C):
\begin{itemize}
\item Embeddings were generated from each newly detected face (starting from 0.5 seconds into the utterance, one frame per second).
\item We computed cosine similarity scores with a threshold of 0.3 to match each face to the speaker's reference embedding.
\item If multiple faces were detected, we selected the one with the highest similarity. If no face passed the threshold, we applied an offset search (looking slightly before or after the initial frame) to find a valid match.
\end{itemize}

This step further filtered out utterances where no suitable face was found. The resulting subset of MELD contains utterances where all three modalities (text, audio, and video) were successfully aligned and the correct speaker's face was detected.

\noindent\textbf{2. Face Extraction for IEMOCAP.}

For the IEMOCAP dataset, face extraction was simpler due to fewer speakers (10 actors across 5 sessions). We applied a similar methodology using DeepFace, YOLOv8, and Facenet-512 with a cosine similarity threshold of 0.3, starting from 0.5 seconds into each utterance. Any utterances without a detected face were discarded. After this process, we retained 7,309 utterances from IEMOCAP that included all three modalities.

\noindent\textbf{3. Audio Channel Selection for MELD.}

MELD's original audio was recorded using a microphone array, resulting in multiple audio channels per utterance. For consistency and quality:
\begin{itemize}
\item From the 6-channel recordings, we selected the third channel.
\item For the 2-channel recordings, we chose the channel with the highest energy level.
\end{itemize}

This ensured a standardized and relatively clean audio input across all utterances. In the end, our experiments relied on subsets of MELD and IEMOCAP where all three modalities were successfully captured and passed our quality control processes. Table~\ref{tab:dataset_stats} summarizes the dataset composition before and after quality control.

\begin{table}[htbp]
\caption{MELD and IEMOCAP Composition Before and After Quality Control}
\label{tab:dataset_stats}
\begin{center}
\begin{tabular}{|l|c|c|c|}
\hline
\textbf{Dataset} & \textbf{Split} & \textbf{Original} & \textbf{Verified} \\
\hline
\multirow{4}{*}{MELD} 
& Train & 9,989 & 7,443 \\
& Dev & 1,109 & 835 \\
& Test & 2,610 & 1,966 \\
\cline{2-4}
& Total & 13,708 & 10,244 \\
\hline
IEMOCAP & Total & 10,039 & 7,309 \\
\hline
\end{tabular}
\end{center}
\end{table}

\section{Results}

\subsection{Overall Performance}
Table~\ref{tab:results} presents emotion recognition performance across modality configurations. On MELD, the trimodal system (T+V+A) achieves 64.8\% accuracy and 64.3\% weighted F1-score. Text emerges as the strongest unimodal baseline at 58.1\%, substantially outperforming vision at 42.3\%. The poor performance of visual features indicates  IEMOCAP shows similar patterns with trimodal fusion reaching 74.3\% accuracy. This suggest there might be temporal misalignment between the emotional peaks for the facial expressions and the extracted frames in our pipelines.

\begin{table}[htbp]
\caption{Emotion Recognition Performance with final fusion by Modality Configuration}
\label{tab:results}
\begin{center}
\begin{tabular}{|c|c|c|c|}
\hline
\textbf{Dataset} & \textbf{Modalities} & \textbf{Accuracy} & \textbf{W-F1} \\
\hline
\multirow{7}{*}{MELD} 
& T & 58.1\% & 57.9\% \\
& V & 42.3\% & 41.8\% \\
& T+A & 62.5\% & 62.1\% \\
& T+V & 61.2\% & 60.9\% \\
& V+A & 48.7\% & 48.2\% \\
& \textbf{T+V+A} & \textbf{64.8\%} & \textbf{64.3\%} \\
\hline
\multirow{7}{*}{IEMOCAP} 
& T & 66.2\% & 66.0\% \\
& V & 52.1\% & 51.7\% \\
& T+A & 72.9\% & 72.6\% \\
& T+V & 71.6\% & 71.3\% \\
& V+A & 58.3\% & 57.9\% \\
& \textbf{T+V+A} & \textbf{74.3\%} & \textbf{74.1\%} \\
\hline
\end{tabular}
\end{center}
\end{table}

\subsection{Class-wise Analysis}
\begin{figure}[htbp]
\centerline{\includegraphics[width=\columnwidth]{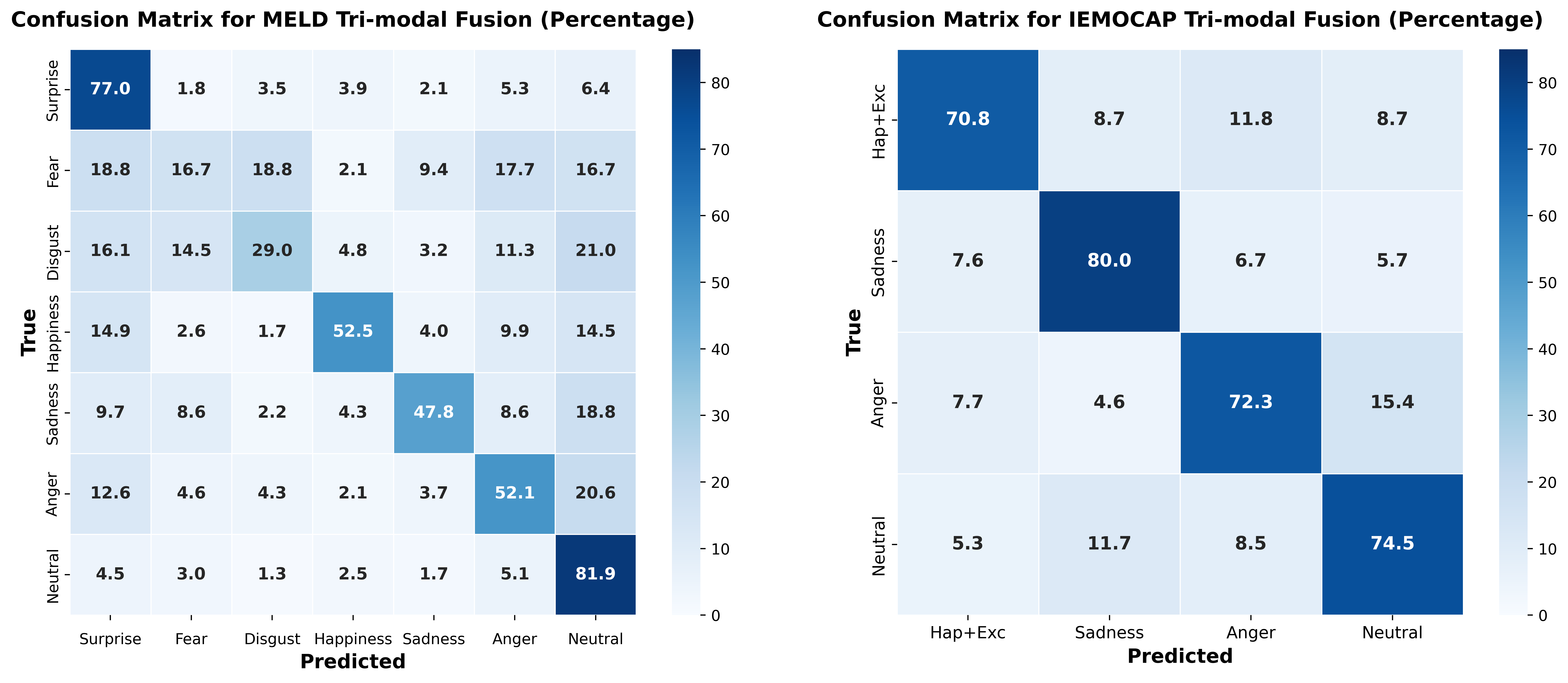}}
\caption{Confusion matrix for trimodal fusion on MELD test set (7 classes) and IEMOCAP test set (4 classes).}
\label{fig:meld-iemocap-confusion}
\end{figure}

Figure~\ref{fig:meld-iemocap-confusion} displays confusion matrices for trimodal fusion. On MELD, neutral achieves 81.9\% accuracy while fear (16.7\%) and disgust (29.0\%) remain problematic. Fear shows substantial confusion with disgust (18.8\%) and anger (17.7\%), suggesting these emotions share overlapping multimodal patterns in conversational contexts. IEMOCAP's four-class problem yields more balanced performance: neutral (74.5\%), sadness (80.0\%), anger (72.3\%), and happy+excited (70.8\%).

\section{Discussion}
Our choice to leverage speaker and face recognition models for emotion transfer departs from conventional approaches, grounded in the hypothesis that identity-preserving representations inherently capture emotion-relevant patterns. Speaker recognition models encode prosodic variations and voice quality while face recognition systems learn subtle facial muscle configurations—both crucial for emotion expression. Similarly, we select MPNet-v2 as a sentence encoder that combines contrastive learning, masked and permuted language modeling, and can be fine-tuned on textual emotion corpora and then used to extract emotion-aware embeddings for conversational datasets such as MELD and IEMOCAP.

The filtered utterances from benchmark data like MELD reveal systematic quality issues, with audio-text misalignment representing most of removed MELD samples. These misalignments could create spurious training signals where models learn incorrect cross-modal associations. The dominance of text (58.1\%) versus poor visual performance (42.3\%) in MELD challenges multimodal fusion assumptions, stemming from compressed TV footage quality, subtle conversational expressions, and temporal misalignment between emotional peaks and extracted frames. The limited fusion improvement (6.7\% over text-only) questions whether current datasets truly require multimodal processing.

This work faces several limitations. We implement quality control without extensive ablation or statistical significance testing, potentially missing systematic biases in our filtering process. The distribution of removed samples across speakers, emotions, and dialogue positions remains unanalyzed. Most critically, our use of fixed speaker embeddings fails to capture temporal dynamics inherent in emotional speech—speaker embeddings provide static representations while emotions evolve over time within utterances.

Future work should address these limitations through several directions. We plan to incorporate pre-trained speech recognition models such as wav2vec 2.0 \cite{r-m:baevski-2020} to enhance the audio component, as these models provide frame-level features that capture temporal emotional evolution unlike static speaker embeddings. This temporal richness would enable better utilization of MAMBA's state-space modeling capabilities, which currently process fixed-dimensional embeddings rather than exploiting sequential speech dynamics. Additionally, developing quality-aware training methods that utilize imperfect samples rather than discarding them could recover the filtered data. We also aim to explore self-supervised approaches learning from unlabeled conversational data and investigate continuous emotion dimensions as alternatives to problematic categorical schemes. Creating new datasets with rigorous quality control during acquisition, rather than post-hoc filtering, remains essential for advancing the field.

\section{Conclusion}
We implement a systematic quality control pipeline for MERC datasets that addresses multiple problematic scenarios—speaker mislabeling, audio-text desynchronization, and face detection failures—ensuring proper alignment across all three modalities. Our approach leveraging identity-based transfer learning achieves 64.8\% accuracy on MELD and 74.3\% on IEMOCAP. However, significant challenges remain: sparse emotions achieve near-random accuracy, and fusion provides limited improvements over text-only baselines. These results emphasize that progress in conversational emotion recognition requires addressing fundamental data quality issues alongside architectural innovations. The field must move beyond incremental model improvements to confront the systematic data problems that limit current approaches, prioritizing the creation of properly aligned, high-quality multimodal datasets and developing methods robust to the inherent ambiguity of emotional expression in natural conversation.

\end{document}